\documentclass[twocolumn,aps,prd,floatfix,preprintnumbers,a4paper,nofootinbib,superscriptaddress,10pt]{revtex4-2}

\usepackage{amssymb,amsmath,verbatim,mathtools,needspace,enumitem,etoolbox,graphicx,microtype}
\usepackage{multirow}
\usepackage[dvipsnames, usenames]{xcolor}
\usepackage{bigints}

\definecolor{linkcolor}{rgb}{0.0,0.4,0.4}
\definecolor{citecolor}{rgb}{.7,.3,.5}
\usepackage[colorlinks=true, linkcolor=linkcolor, citecolor=citecolor, pdfusetitle]{hyperref}

\usepackage[all]{hypcap}
\usepackage[T1]{fontenc}
\usepackage[utf8]{inputenc}
\usepackage{graphicx,amssymb,amsmath,amsthm,amsfonts,epsfig,times,natbib}
\usepackage[normalem]{ulem}
\usepackage{multirow}
\usepackage{import}
\usepackage[titletoc,title]{appendix}

\usepackage{graphicx}				
\usepackage{placeins}
\usepackage{subcaption}
\usepackage{ragged2e} 
\usepackage{array}
\DeclareCaptionFormat{myformat}{\justifying#1#2#3}
\usepackage{verbatim}
\usepackage{rotating}
\usepackage{comment}
\usepackage{txfonts}

\usepackage{epstopdf}
\usepackage{tensor}
\usepackage{amsbsy}
\usepackage{bm,url}
\usepackage{float}
\usepackage{dcolumn}
\usepackage{latexsym}
\usepackage{rotating}
\usepackage{enumerate}
\newcommand{\msun}{\ensuremath{M_\odot}}

\def\be{\begin{equation}}
\def\ee{\end{equation}}
\def\bea{\begin{eqnarray}}
\def\eea{\end{eqnarray}}

\newcommand{\ba}{\begin{align}}
\newcommand{\ea}{\end{align}}

\newcommand{\deco}{\texttt{PhenomDECO}}
\newcommand{\dtap}{\texttt{PhenomDTaper}}
\newcommand{\bbh}{ \texttt{PhenomD}}

\newcommand{\AEI}{Max Planck Institute for Gravitational Physics (Albert Einstein Institute),
Callinstrasse 38, D-30167 Hannover, Germany}
\newcommand{\Leibniz}{Leibniz University Hannover, 30167 Hannover, Germany}
\newcommand{\Cardiff}{School of Physics and Astronomy, Cardiff University, Cardiff, CF24 3AA, United Kingdom}

\begin{document}
\title{On the Identification of Exotic Compact Binaries with Gravitational Waves: a Phenomenological approach}
\author{Shrobana Ghosh}
\affiliation{\AEI}
\affiliation{\Leibniz}

\author{Mark Hannam} 
\affiliation{\Cardiff}

\begin{abstract}
Gravitational wave (GW) astronomy has been hailed as a gateway to discovering unexpected phenomena in the universe. Over the last decade there have been close to one hundred GW observations of compact-binary mergers.  While these signals are largely consistent with mergers of binary black holes, binary neutron stars, or black hole-neutron star systems, some events suggest the intriguing possibility of binaries involving exotic compact objects (ECOs). Identifying and characterising an ECO merger would require accurate ECO waveform models. Using large numbers of numerical relativity simulations to develop customised models for ECO mergers akin to those used for binary black holes, would be not only computationally expensive but also challenging due to the limited understanding of the underlying physics. Alternatively, key physical imprints of the ECO on the inspiral or merger could in principle be incorporated phenomenologically into waveform models, sufficient to quantify generic properties. In this work we present a first application of this idea to assess the detectability and distinguishability of ECO mergers, and we propose a phenomenological approach that can iteratively incorporate features of ECO mergers, laying the groundwork for an effective exotic compact object identifier in compact binary coalescences. Using Bayesian parameter estimation on the data for the GW event GW150914, we find the inferred compactness to be consistent with that expected for black holes, within this framework. The efficacy of the identifier can be refined by adding information from numerical relativity simulations involving fundamental fields. Conversely, such an identifier framework can help focus future numerical relativity and modeling efforts for exotic objects.
\end{abstract}

\maketitle

\section{Introduction}
Over the last decade the GW detectors~\cite{ LIGOScientific:2014pky, VIRGO:2014yos, PhysRevD.88.043007} have recorded over a hundred observations of compact binary coalescences (CBCs) \cite{Abbott_2023, Nitz_2023, PhysRevD.101.083030, PhysRevD.100.023007}. CBCs are a powerful source of GWs due to their rapidly changing mass quadrupole, and a promising target for new sources would be mergers of compact objects that are \emph{not} black holes or neutron stars. Of the hundred detections, GW170817 \cite{PhysRevLett.119.161101} --  the coalescence of two neutron stars -- is arguably the only event that allowed for confident source identification due to the multimessenger nature of the detection \cite{Abbott_2017_multi}.  While the signal for all of the events are consistent with binary black hole (BBH), binary neutron star (BNS) or black hole-neutron star (BH-NS) mergers, the possibility of the binary comprising exotic compact objects (ECOs) remains. In fact some of the more interesting events -- GW190521, GW190814, GW190425 -- have spurred investigation of various scenarios \cite{clesse2021gw190425, PhysRevLett.126.081101, PhysRevLett.125.261105, aurrekoetxea2023revisiting} due to the observation of compact objects in the lower or upper mass gaps.


A key difficulty in current GW searches is that the observation of a source is largely dependent on having a signal model with which to uncover the signal in the data stream, due to the signals being typically weak compared to detector noise. While unmodeled searches that look for coherent power in the data stream can be useful in looking for unknown signals, they have limitations in terms of length of signal and also in accurately inferring source properties. One approach to modelling exotic-compact-object (ECO) mergers would be to follow the procedure used for BHs and NSs: combine numerical-relativty (NR) simulations of these mergers with approximate analytical predictions of the inspiral signal.
Such an approach is computationally extremely expensive (note that complete models of generic BH and NS mergers signal do not yet exist), and is in principle limited to the class of object under consideration. An alternative approach is to consider the phenomenology of various classes of ECO, and to make approximate phenomenological models that allow us to study how easy it is to detect and identify these objects, and which properties we may be able to measure. By their nature such models may not be as discriminating as those constructed from accurate NR simulations, but they have the advantage of being relatively easy to construct, and may provide a guide as to which classes of objects are most promising to search for, or where detailed NR studies will be most useful. One could also imagine a hybrid approach where a small number of NR simulations are used to gain a general picture of the phenomenology of different physical effects, without the need for a fine sampling of thousands of simulations across a high-dimensional parameter space.
 
To remain as model-agnostic as possible, key physical imprints of exotic compact objects (ECOs) during the inspiral or merger regime could be incorporated phenomenologically into waveform models to capture and quantify their generic properties, building upon point particle approximants. We present a first application of this approach and assess their applicability to ECO mergers. We further propose a framework capable of iteratively integrating ECO-specific features and laying the foundation for an effective method to identify exotic compact objects in compact binary coalescences.

{ \ }
ECOs have been proposed with various motivations, as compact objects that are not plagued with pathological singularities or remedies to the ``information-loss paradox", among others~\cite{Maggio_2021}. Notably, a self-gravitating configuration of a complex bosonic field has been considered in great detail and is well known in the literature today as a \textit{boson star} \cite{Kaup:1968zz}. Although the mechanism that enables such configurations to avoid gravitational collapse is unclear, unlike in neutron stars, stable solutions exist in the literature (see \cite{Liebling_2023} for examples). A popular notion is to consider boson stars as a Bose-Einstein condensate on an astrophysical scale; this allows interpreting the system as a quantum macrostate and then a gravitational collapse is averted by the Heisenberg uncertainty principle~\cite{Giudice_2016}.

The morphology of the GW signal from an ECO binary is expected to differ from that of a BBH primarily due to differences in compactness of the objects and tidal interactions. This suggests that a careful consideration of the contribution to the phase of the binary due to tidal deformation by a companion, tidal heating, spin-induced quadrupole moments or self-interactions, in case of boson stars, can distinguish ECO binaries from BBHs. In fact the distinguishability of ECO binaries from BBHs has been explored in some detail by perturbative modifications to the inspiral phase~\cite{Wade:2013hoa, Cardoso:2017cfl, Sennett:2017etc, Krishnendu:2017shb, Krishnendu:2019tjp, krishnendu2025testingnaturecompactobjects, Johnson-McDaniel_2020, Datta:2019epe, Datta:2020gem, Porto:2016zng, PhysRevLett.120.081101,Narikawa:2021pak, Siemonsen:2023hko}. More recently a templated search with an inspiral only model for compact objects was conducted on the publicly available data from the LIGO-Virgo-KAGRA (LVK) collaboration~\cite{Chia:2023tle} by incorporating effects of large tidal deformabilities into the phase modulation and obtaining model independent constraints on the Love numbers. Additionally, the remnant of a merger of an ECO binary may produce an object of similar origin, or a BH, making ringdown tests a good method for distinguishability have also been considered~\cite{PhysRevD.88.064046, PhysRevLett.116.171101, PhysRevD.93.064053, Yunes:2016jcc}.
\begin{figure}[h]
\includegraphics[width=0.49\textwidth]{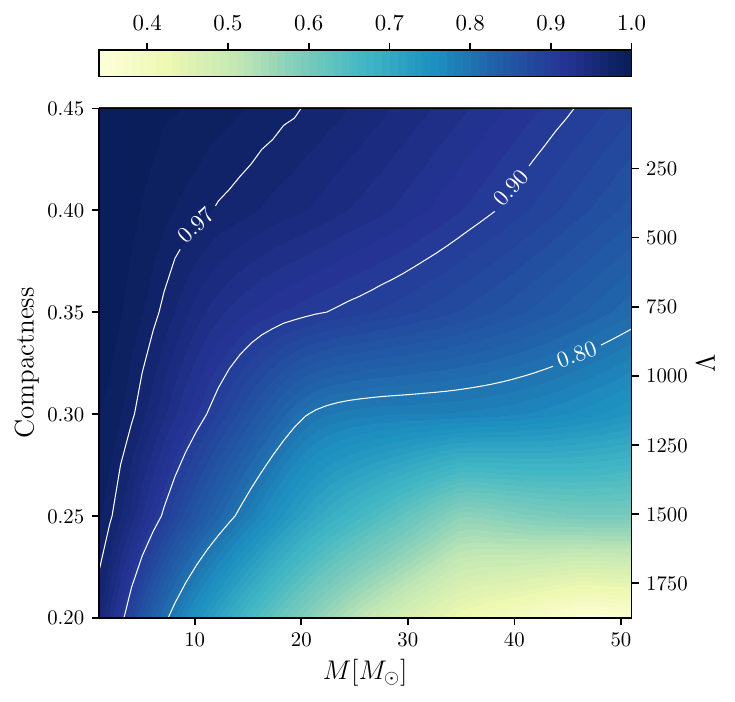}
\caption{Match contours shown for PhenomD\_NRTidalv2 and PhenomD for a range of compactness (tidal deformabilities $\Lambda_1=\Lambda_2=\Lambda$, $y$-axis) and total mass ($x$-axis). 
We expect tidal effects to have minimal impact above the $\mathcal{M}=0.97$ contour, and to be important below the $\mathcal{M}=0.8$ contour. 
}
\label{fig:tidalphase}
\end{figure}

Under the assumption of a polytropic equation-of-state to describe the pressure-density relationship of the constituent of an ECO, specifically a boson star, it is possible to infer the tidal parameters~\cite{Johnson-McDaniel_2020}. Ref.~\cite{Johnson-McDaniel_2020} applies contributions from tidal deformation due to the companion as well as from spin-induced quadrupole moments consistently in the inspiral phase of the binary. Noting that compact objects of lower compactness will merge sooner than BHs, they compute a contact frequency that is self-consistent with their choice of equation-of-state. However, their treatment inherently assumes strongly self-interacting bosons and therefore excludes other ECOs. Furthermore, the polytropic assumption for the equation of state allows probes of only a limited range in compactness of the boson star. 

Aside from phase modulation due to tidal effects, the post-contact part of the signal from an ECO binary may differ significantly from that of a BBH merger. To understand the rich morphology of the signal in this regime numerical simulations are needed. Apart from the usual challenge of computational costs, simulating exotic binaries presents the additional challenge of modelling the matter comprising the ECOs. For the specific case of boson star binaries, simulations of head-on collisions~\cite{Palenzuela:2006wp, Dietrich:2018bvi, Clough:2018exo}, orbiting boson stars~\cite{Palenzuela:2007dm, Bezares2017, Palenzula2017} and the latest breakthrough of full numerical simulations of boson star binaries~\cite{Evstafyeva:2024qvp} have shed light on this class of ECO mergers. Ref.~\cite{Toubiana_2021} uses a clever workaround inspired by~\cite{Takami}, where they model the binary in the last stages as two point particles connected by a spring. The restorative force of the spring allows the exotic cores to bounce off each other and the whole set up is embedded in a disk that models the disrupted material. Based on the success of this toy model in reproducing some of the key features of the post merger phase for neutron star binaries it is reasonable to expect similar performance here, however careful consideration of the properties of the disk may be needed depending on the nature of the exotic object (eg. for \textit{fuzzballs}). It is worth noting that with a refined version of this toy model it may be possible to make independent measures of equation-of-state parameters from inspiral stage ($\tilde\Lambda$) and post-merger stage (\textit{spring constant} $k$) and increase confidence on the equation-of-state inferred from neutron star binaries.

Based on the discussion above, the possibility of missing ECO binaries in current matched-filtering searches seems significant, as is evident from the poor matches between BBH and BNS waveforms in Fig.~\ref{fig:tidalphase}, particularly for lower compactness and masses roughly above 10 $M_\odot$. However, as is shown in recent simulations of inspiral-merger-ringdown (IMR) of boson star binaries~\cite{Evstafyeva:2024qvp} the tidal deviation to the phase may be insignificant in some cases. Furthermore, figure~\ref{fig:tidalphase} also depicts that for moderately low compactness a BBH waveform would be just as good to detect very low mass BNSs, this may also be true for exotic objects. As such, to the best of our knowledge, the detectability of ECO binaries with BBH waveforms has not been explored systematically across the total-mass--mass-ratio parameter space that current GW detectors are most sensitive to. 

To make an approximate phenomenological model based on universal features of the inspiral, we note that compact objects larger than black holes can be expected to merge at a larger separation, and so the inspiral will terminate at a lower frequency. Based on this simple observation we modify the amplitude of a standard  BBH waveform model, ~\bbh, to terminate at a lower frequency, parameterised by the object's compactness. For this first illustrative exercise we do not attempt to model any other physics, e.g., tidal deformations during the inspiral, or the nature of the merged object; we make only one change to the waveform, and assess its impact. Furthermore, a distinguishability criterion for putative BBH detections has not been defined to date. Our aim in this work is to explore the plausible range of physical parameters for ECOs, assess their detectability by current generation GW detectors and propose an ``ECO identifier'' that can discriminate between a BBH and non-BBH hypothesis, without loss of generality. We note here that a similar early end of inspiral has been noted for neutron star-black hole binaries~\cite{Lackey:2011vz} or sub-solar mass binaries~\cite{Crescimbeni:2024cwh}, as well as for exotic binaries~\cite{Chia:2022rwc,Chia:2023tle}, based on an earlier peak of the phase derivative due to the inclusion of tidal terms, and have been used in state-of-the-art CBC waveforms that include matter effects~\cite{Dietrich:2017aum,Thompson:2020nei}. Thus, our amplitude modification does in fact include the most dominant impact of phase modulation due to tidal effects, without making specific assumptions about the underlying physics. 


We outline and motivate ad-hoc modifications to BBH waveform used in this study in Sec.~\ref{sec:outline}. In Sec.~\ref{sec:detectablity} we discuss detectability of ECOs by BBH waveforms and biases in recovery for detectable systems. We propose an identifier for ECOs in a binary in Sec.~\ref{sec:compactness} based on the modified BBH model.

\section{Modelling exotic mergers}
\label{sec:outline}
\subsection{Contact frequency}
\label{sub:contact}
The separation or frequency at which transition from inspiral to merger regime for a BBH coalescence happens is hard to determine. It is standard to assume that the final plunge begins at separations just below the innermost stable circular orbit ($r_{\rm{ISCO}}$) which is at $r^{BH}_{\rm{ISCO}} =6m_{\rm{BH}}$ for a Schwarzschild BH of mass $m_{\rm{BH}}$~\cite{Bardeen:1972}.  Noting that compactness of an object can be defined as $C\coloneqq \frac{m_{\rm{BH}}}{r_H}$, where $r_H$ is the black hole hoirzon and that $C=\frac{1}{2}$ for a Schwarzschild BH, $r_{\rm{ISCO}}$ can be rewritten as $\frac{3m_{\rm{BH}}}{C}$~\cite{Giudice:2016zpa}. In fact relativistic calculations including tidal effects for neutron stars corroborates this simplistic relationship of $r_{\rm{ISCO}}$ with compactness ~\cite{Dong1996}, of course $r_H$ would correspond to the size of the exotic object. 

In the inspiral the point particle approximation used for BBHs captures the dominant features of the waveform for ECOs just as well. But we expect the inspiral to end earlier due to the larger size of the objects. For an accurate estimation of separation at ``contact", corrections due to tidal effects need to be incorporated~\cite{PhysRevD.81.084016}, but at leading order it can be written as 
\be
\label{eq:contact}
R_c = R_1 + R_2,
\ee
where $R_i$ denotes the size of objects $i= 1,  2$. To make a theory-agnostic choice for separation at contact, we can also generalise the definition of ISCO to non-rotating compact objects as
\be
\label{eq:risco_eco}
r^{\rm{ECO}}_{\rm{ISCO}} = \frac{3m_{\rm{ECO}}}{C} \; .
\ee
This allows us to treat ECOs as horizonless compact objects with an outer boundary and without any assumption on their formation mechanism. We only assume that the two objects are expected to make contact when the separation falls below ISCO.  

A representative comparison of $R_c$ for $C=0.25$ (corresponds to an ECO twice as large as a BH) shows that Eq.~(\ref{eq:risco_eco}) in fact gives a larger separation than Eq.~(\ref{eq:contact})
\begin{subequations}
\begin{align}
R_c =& 4m+4m = 8m \label{subeq1}\\
r^{\rm{ECO}}_{\rm{ISCO}} =& \frac{3m}{C} = 12m \label{subeq2}
\end{align}
\end{subequations}
with $m_1 = m_2 = m$ in an equal mass ECO where the suffix ``ECO'' has been dropped for brevity. Including tidal corrections would lead to larger values of $R_c$ (c.f Eq. 2 in Ref.~\cite{Johnson-McDaniel_2020} and descriptive text). Therefore, actual contact should occur somewhere between equations~(\ref{subeq1}) and (\ref{subeq2}). 

The frequency of the GW signal emitted close to plunge (ISCO) for a BBH using Kepler's law is
\be
\label{eq:fisco_bh}
f^{BH}_{\rm{ISCO}} = \frac{1}{6^{3/2}\pi \bar{M}} \,;
\ee
we use $\bar{M}$ to indicate that either the primary mass or the total mass could be used for unequal mass ECOs. Using equations~(\ref{eq:risco_eco}) and~(\ref{eq:fisco_bh}) the \textit{contact frequency} for ECOs can be parametrized by their compactness as 
\be
f^{\rm{ECO}}_{\rm{ISCO}} = (2C)^{3/2} f^{BBH}_{\rm{ISCO}} \; .
\label{eq:ECOf}
\ee
Note that using Eq.~(\ref{subeq1}) would yield a higher frequency for the inspiral and that would correspond to a compactness of $C=0.375$ in Eq.~(\ref{subeq2}). In other words using the ISCO for separation at contact will produce conservative estimates for detectability of these signals. On the other hand, the leading order relativistic corrections to Kepler's law being negative (c.f. Eq.~228 in Ref.~\cite{Blanchet:2013haa}) would lead to a lower ISCO frequency. So, a full model for analysing ECO binaries should account for that. 

\subsection{Tapered model}
\begin{figure}[h]
\includegraphics[width=0.49\textwidth]{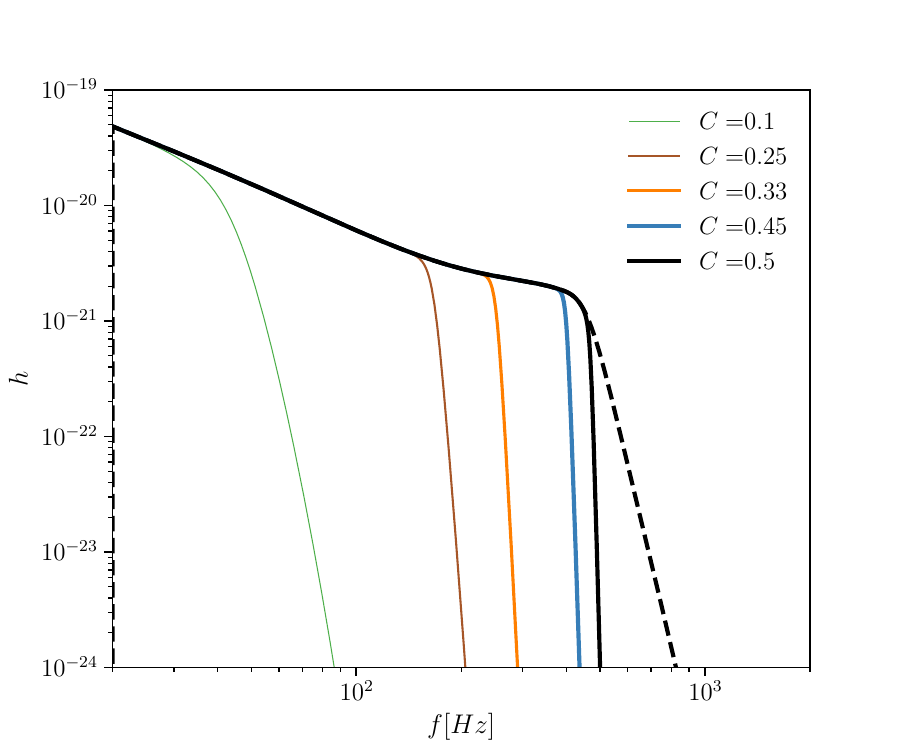}
\caption{The frequency domain amplitude  from ~\bbh~ for an equal-mass BBH of $M_{\rm{tot}} = 40 M_\odot$ is shown in dashed black. Modified amplitude for ECO binaries of the same total mass with compactness $C= 0.1, 0.25, 0.33, 0.45$ are shown by lines of increasing thickness. Even at $C=0.5$ (solid black line), the BBH amplitude is not fully reproduced.}
\label{fig:dtapamp}
\end{figure}

Since we expect the inspiral for ECOs to end earlier than for BBHs, evidently lower compactness would lead to lower contact frequency. With this in mind, we modify the amplitude of the aligned-spin~\bbh~approximant~\cite{Khan:2015jqa,Husa:2015iqa} such that it tapers off at the contact frequency corresponding to the compactness of the ECOs as obtained from eq.~(\ref{eq:ECOf}). However, noting that we are using a full IMR approximant, and that contact is closer to merger than to the ISCO, especially for higher masses (see table~\ref{tab:frequencies} for example), we use the BBH ringdown frequency to parameterize the tapering frequency by compactness, 
\be
A^{\rm{ECO}}=\frac{A^{\rm{BBH}}}{2}\Bigg[1-\tanh\frac{\Big(f-(2C)^{3/2}f_{\rm{RD}}\Big) }{f_D}\Bigg] .
\ee
\begin{table}[hb]
\begin{center}
\begin{tabular}{ | m{1cm} | m{1cm} | m{1cm}| m{1.5cm} | m{1.5cm} | m{1.3cm} | } 
  \hline
  M & $m_1$ & $m_2$& $f_{\rm{ISCO}}^{m_1}$ & $f_{\rm{ISCO}}^{m_1+m_2}$ & $f_{\rm{RD}}$ \\
    \hline
  $150\msun$ & $75\msun$ & $75\msun$ & $58.66$ Hz & $29.33$ Hz & 125.11 Hz \\
  \hline
  $150\msun$ & $120\msun$ & $30\msun$ & $36.66$ Hz & $29.33$ Hz & 106.2 Hz \\
  \hline
\end{tabular}
\caption{Different considerations of the frequency to be parameterized by compctness for producing~\dtap~; we use  total mass for $f_{\rm{ISCO}}^{m_1+m_2}$ and primary mass for $f_{\rm{ISCO}}^{m_1}$. This shows that the ISCO would significantly underestimate the frequency close to contact.}
\label{tab:frequencies}
\end{center}
\end{table}
\\
We use a $tanh$ window to ensure a smooth fall off and the window width, $f_D$, gives us some flexibility in adjusting the slope of the fall. 
We choose a conservative window width of $f_D=10$ Hz, as shown in figure.~\ref{fig:dtapamp} instead of a , for inspecting detectability of ECOs by inspiral-merger-ringdow (IMR) waveforms.
%
As seen in fig.~\ref{fig:tidalphase} for neutron stars, it is very likely that for high tidal deformation (low compactness) the phase will be so different that the ECO signals will be completely missed by a template that does not account for the dephasing. However, for high compactness and certainly for low masses ($\lesssim 10\msun$) the dephasing will not be significantly impactful. Additionally, for very low compactness the amplitude cut-off will be at a much lower frequency compared to the BBH counterpart, as can be seen for $C=0.1$ in fig.~\ref{fig:dtapamp}. Hence it will  be the dominant effect on the signal. Furthermore, a detectability study with specific dephasing tuned to a particular kind of ECO would only be insightgul about that family of ECO. In other words, while using BBH waveforms without accounting for phase modulations or post-contact signal may overestimate an ECO's detection prospects, it preserves the general and dominant features of the chirp in compact binaries. Concerns on overestimation of detectability are easily dispelled by noting that the purpose of this study is not to propose using BBH waveforms to look for ECOs, but to identify regions where our current searches are possibly missing these signals and regions where we may already be seeing them and mischaracterising them as BBH mergers. Furthermore, a qualitative assessment based on universal features of an inspiral (e.g., a physically motivated modification to the amplitude envelope) can be useful to identify the generic ECO parameter space that needs to be focussed on when making models tuned to specific ECO properties. 

\subsection{\deco}

\dtap~drops off abruptly at ``contact''; evidently it cannot produce the BBH limit even for a compactness of 0.5. To help assess the impact of the details of our modelling approach, a model that smoothly approaches the BBH limit as $C \rightarrow 0.5$ is more useful. In this section we describe our \textit{first attempt towards refashioning \rm{\texttt{PhenomD}} into a phenomenological waveform that admits compactness} as an intrinsic parameter. Compactness encodes the mass-radius information of an object and is a good criteria to differentiate between strongly gravitating objects, such as black holes, neutron stars or anything in between. 
%
\begin{figure}
\includegraphics[width=0.49\textwidth]{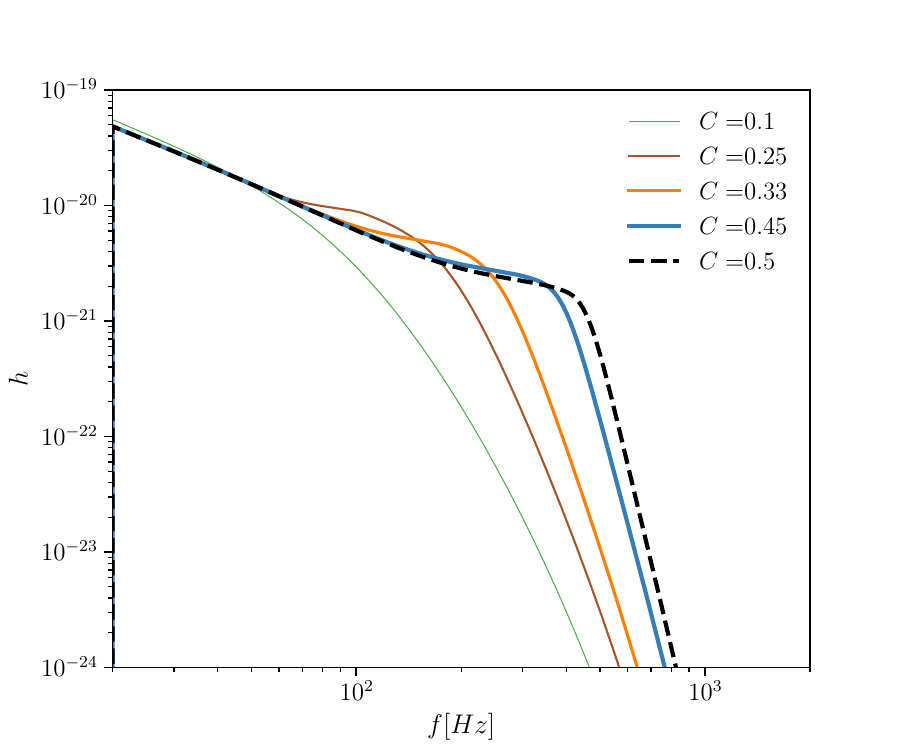}
\caption{Same as fig.~\ref{fig:dtapamp} with the amplitude parameterised by compactness. We show the amplitude for four different values of the compactness parameter, going from lower to higher values represented by progressively thicker lines. We also show that for C=0.5 ~\deco~ reproduces the BBH amplitude as shown by the black dashed line.}
\label{fig:decoamp}
\end{figure}

We provide here a quick recapitulation of the construction of the \texttt{PhenomD} approximant and point the readers to the original papers~\cite{Khan:2015jqa,Husa:2015iqa} for details. Hybrids constructed from post-Newtonian (PN) and numerical relativity waveforms were fitted to three different ans\"atze for the inspiral, the merger-ringdown and an intermediate region. The final approximant then consists of the ans\"atze and the parameterized fits of the fitting coefficients. This means that the model is able to smoothly transition from one region to the other at specific frequencies. These transition frequencies (reproduced here from Table II on Ref.~\cite{Khan:2015jqa,Husa:2015iqa} only for readability)
\begin{align*}
f_1 &= 0.014\\
f_3 &= \Bigg |f_{\rm{RD}} + \frac{f_{\rm{damp}}\gamma_3 \bigg( \sqrt{1-\gamma_2^2}-1\bigg)}{\gamma_2} \Bigg |
\end{align*}
were determined by an optimization of the fits to the PN-NR hybridised waveforms. In~\deco~ we rescale both these frequencies, like $f_{\rm{ISCO}}$ in~\dtap, by the compactness parameter 
\begin{align*}
f_1^{\rm{ECO}} &= (2C)^{\frac{3}{2}}f_1\\
f_{\rm{RD}}^{\rm{ECO}} &= (2C)^{\frac{3}{2}}f_{\rm{RD}}^{\rm{ECO}}.
\end{align*}
The amplitude is rescaled such that the amplitude of the intermediate region smoothly transitions to that of the merger-ringdown region at this new value of $f_3$. In the absence of a similar intuition for the phase we have left the phase untouched -- all three waveforms (~\bbh~,~\dtap~,~\deco~) have the same phase -- and intend to explore this in future work. 

The purpose of incorporating the compactness in two different ways is to allow us to assess the robustness of results. Comparing the amplitude profiles of~\dtap~ and ~\deco~ it is easy to note that for the same parameters ~\deco~ has slightly more power in the signal. Consistent with that we find ~\deco~ just leads to slightly more optimistic estimates on detection and does not change conclusions reached by the ~\dtap~ study. Therefore, we only report the results of~\dtap~ on detectability. On the other hand, introducing the compactness parameter in this way allows generating BBH waveforms for $C=0.5$, as shown by the dashed black line on Figure~\ref{fig:decoamp}. This is particularly interesting in the context of bayesian inference as it allows us to sample over compactness parameter to generate an ECO binary (as per our definition) or a BBH enabling us to use standard data analysis techniques to verify if the compactness of the merging objects are indeed consistent with that of BHs. Interestingly deviation from $C=0.5$ could also be indicative of significant spin on one or both of the components BHs. We explore compactness estimates obtained with~\deco~ in details in sec.~\ref{sec:compactness}.
 %
\section{Detecting exotic mergers}
\label{sec:detectablity}
\begin{figure*}[t]
\includegraphics[width=\linewidth]{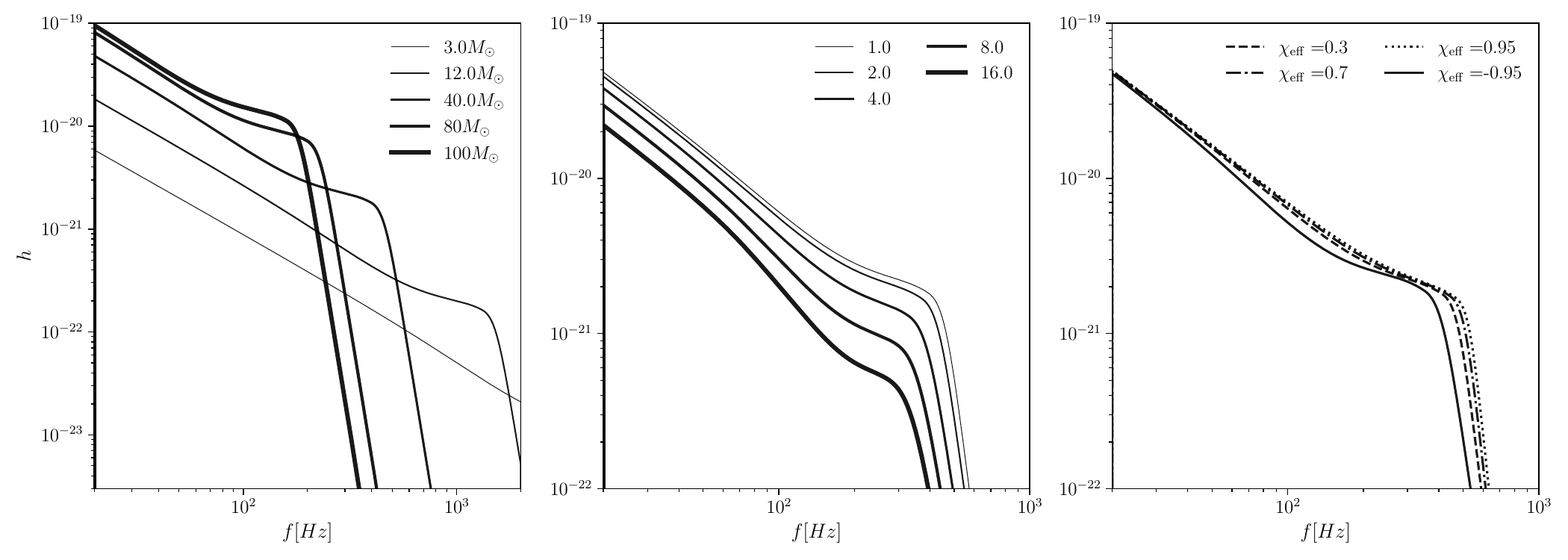}
\caption{Dependence of the frequency-domain amplitude of a BBH signal on the total mass of the binary (left) for an equal mass binary, mass ratio (center) for a total mass of $40 M_\odot$ and effective spin (right) for an equal mass, $M=40 M_\odot$ binary. The amplitude of higher total mass are depicted by lines of progressively higher thickness; similarly, progressively lower mass ratio correspond to lines of progressively higher thickness. For effective spin, we show anti-aligned spin in solid black, while aligned spins are shown by dashed, dashed-dotted and dotted lines for $\chi_{\rm{eff}}= 0.3, 0.7, 0.95$ respectively. It is easy to note that both mass ratio and total mass can interfere with an earlier cut-off of inspiral; the effect of $\chi_{\rm{eff}}$ is much more nuanced and can cause subtle changes in the shape of the amplitude profile.}
\label{fig:ampprofile}
\end{figure*}
Assuming that the models discussed in the previous section are a crude, but physically meaningful, representation of the signal from an exotic binary merger, we now focus on assessing if waveforms used in current GW searches would be able to detect such signals. Waveforms used in CBC searches have undergone iterative improvements and the latest generation models used in inferring source properties now include subtle effects such as precession, impact of higher multipoles, multipole asymmetries etc. While inclusion of these effects are imperative for precise estimates of the source parameters, they have no impact on the detection problem, and it is therefore standard practice to use aligned spin waveforms in searches. Bearing that in mind we will use the aligned-spin waveform ~\bbh~ as the template in the detectability analysis.


%
%
We explore the detectability of ECO mergers with signals generated using ~\dtap~ as well as ~\deco~ but show the results only for ~\dtap~ because it yields the more conservative estimates between the two. This can be easily understood by the sharp amplitude fall off in~\dtap~ that reduces the power in the signal compared to ~\deco~. We will focus on objects whose compactness lies in the range $0.1\leq C < 0.5$; $C=0.25$ corresponds to objects twice as large as a Schwarzschild BH of the same mass. We assume the same compactness for both objects and that they are non-spinning. We will consider Bayesian inference to detect ECOs by~\deco~ in Sec~\ref{sec:compactness}.


\subsection {Detecting ECO mergers with a BBH model}

The dependence of the amplitude of the frequency-domain GW signal from a BBH on the intrinsic parameters of the binary and the luminosity distance to the source are well known. Fig.~\ref{fig:ampprofile} helps to contextualize these quantities for the detectability analysis. Unsurprisingly, the total mass of the system, ratio of the mass of the larger BH to the smaller one and the effective spin parallel to the orbital angular momentum can all mimic an early cut-off of the inspiral. 


We use standard techniques for CBC data analysis to compute the time and phase optimized overlap -- \textit{match}~\cite{Cutler:1994ys,Owen:1995tm} -- between the signal ~\dtap~/~\deco~and templates generated using the waveform model~\bbh~ in the frequency-domain,
\be
\begin{split}
\mathcal{M} \equiv \frac{\langle h_{\rm{PhenomD}} | h_{\rm{ecomodel}} \rangle}{\sqrt{\langle h_{\rm{PhenomD}} | h_{\rm{PhenomD}} \rangle \langle h_{\rm{ecomodel}} | h_{\rm{ecomodel}} \rangle}}
\end{split}
\ee
where,
\begin{equation*}
\langle h_{\rm{PhenomD}} | h_{\rm{ecomodel}} \rangle=  4 \mathfrak{R} \int \frac{h_{\rm{PhenomD}}(f) h_{\rm{ecomodel}}^*(f)}{S_n(f)} df .
\end{equation*}
The match is computed using the implementation in the \texttt{pycbc} library~\cite{alex_nitz_2024_10473621}, using the advanced-LIGO detector sensitivity curve~\cite{adLIGOnoise} with a lower frequency cut-off of 20 Hz, and then we optimize over the intrinsic parameters $(q, M, \chi_{\rm{eff}})$ of the template to obtain the \textit{fitting factor},$\mathfrak{F}$~\cite{Apostolatos}. 
\begin{figure*}[t]
\includegraphics[width=\textwidth]{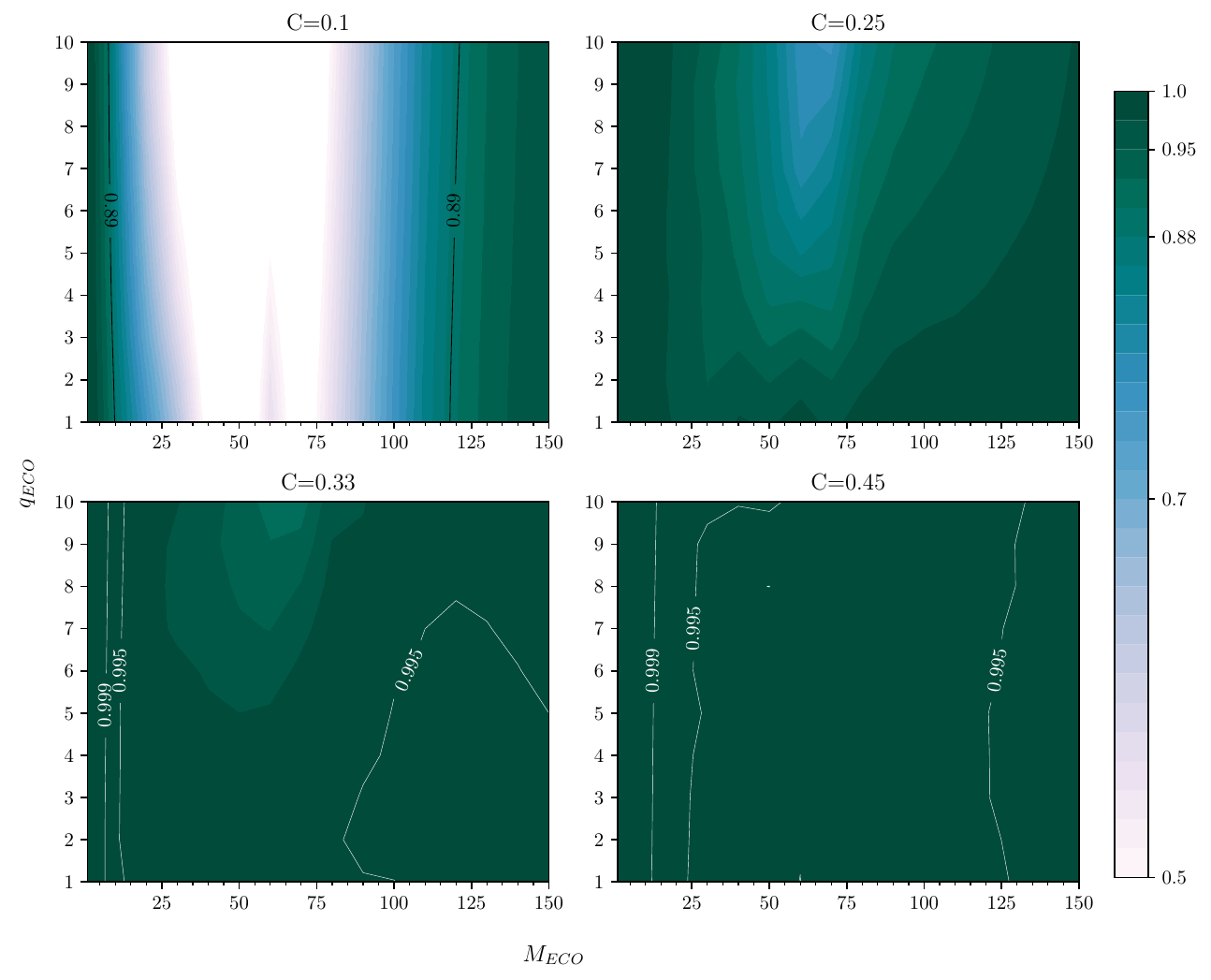}
\caption{Fitting factor contours obtained by optimizing the match between~\dtap~and~\bbh~over the $q-M-\chi_{\rm{eff}}$ parameter space for different ECO total masses (\textit{x-axis}) and mass ratio (\textit{y-axis}); \textit{top left: } $C=0.1$, \textit{top right: } $C=0.25$, \textit{bottom left: } $C=0.33$, \textit{bottom right: } $C=0.45$ and colour bar on right shows the value of the fitting factors. The black contours on the top left panel enclose the region of $q-M$ space for non-spinning ECOs that would be interesting in residual tests, if such ECOs exist in nature.
}
\label{fig:detectability}
\end{figure*}
The fitting factor gives us the fraction of the true signal-to-noise ratio that can be recovered by a waveform model. To be precise, if~\dtap~/~\deco~ represents a true ECO merger that has an SNR $\rho$ in the detector, then a~\bbh~ template with fitting factor $\mathfrak{F}(<1)$ can only recover a SNR of $\mathfrak{F} \rho  $. This means if the detectability threshold in a search is $\rho_{\rm{threshold}}$, a model with $\mathfrak{F}<1$ will only be able to detect signals that have an $SNR=\rho_{\rm{threshold}}/\mathfrak{F} $. Therefore, calculating fitting factor with a waveform model used in standard searches gives an understanding of detectability of ECO mergers with a standard BBH template. Figure~\ref{fig:detectability} shows the fitting factor obtained by optimising over a three-dimensional subspace of the full parameter space that affects the template amplitude i.e., the total mass, mass ratio and effective spin, for different total masses and mass ratio of ECO binaries. As is evident from the plot for a very low compactness ($C = 0.1$) the fitting factors are quite low. 

In general, for any compactness $\mathfrak{F}$ gets worse with increasing total mass, hitting a minimum at $\sim 30-70 M_\odot$ (depending on the compactness) and then starts improving again. This behaviour can be easily understood by taking note of the different panels of Fig.~\ref{fig:ampprofile}. The amplitude drop-off at an earlier frequency is morphologically similar to a higher mass binary. For really low total masses, the signal is in band at most till the late stages of inspiral. Therefore, the amplitude dropping off earlier than when a BBH ringdown sets in has very little impact on the waveform. On the other end of the spectrum, for high masses, very little of the inspiral is ``visible'' to the detectors. So the earlier onset of merger in the signal leads to a small fraction of power being lost in the match computation. Naturally, the mass range for which the full IMR signals lie entirely within a detector's sensitivity bandwidth yield the worst matches. 

For moderately compact objects ($C\gtrsim0.25$) the $\mathfrak{F}$ never falls below $\sim0.9$ as long as the binary is not very highly asymmetric in mass. However, the actual ability to detect in a search is a more nuanced discussion. For instance, a $\mathfrak{F}<0.5$ as can be seen for an ECO binary with $C=0.1$ and $M_{\rm{ECO}}=50\msun$ would not be detectable if the true SNR of the signal is 15, assuming a $\rho_{\rm{threshold}} =8$. However, if the signal has an SNR of 20 or more it would be detected, provided it passes the usual consistency checks done in a BBH search. Nevertheless, fig.~\ref{fig:detectability} gives an indication of the fraction of ECO binaries that would be missed in a population using a BBH search. Noting that the SNR roughly scales as $\rho \propto \frac{1}{D_L}$, $D_L$ being luminosity distance to the source, a waveform that recovers $0.9\rho$ is sensitive to only a distance of $0.9 D_L$ (and a volume of $0.9^3 = 0.73$).



A standard test used in detecting deviations from general relativity (GR) as the underlying theory of gravity by the GW community involves inspecting the residual SNR after subtracting the best matching template from the data. The residual SNR is related to the fitting factor by~\cite{LIGOScientific:2016lio}
\begin{equation}
\rho_{\rm{res}}^2 = (1 - {\mathfrak{F}}^2) \frac{\rho_{\rm{BBH}}^2}{{\mathfrak{F}}^2} ,
\label{eq:residual}
\end{equation}
where $\rho_{\rm{res}}$ is the residual SNR and $\rho_{\rm{BBH}}$ is the maximum SNR that would be recovered by a BBH template. If we assume a residual SNR of 8 would prefer the hypothesis that the residual is consistent with a coherent signal rather than with Gaussian noise, then a trigger with $\rho_{\rm{BBH}} = 16$ may have enough residual SNR if it was generated an by an ECO signal of $\mathfrak{F}\leq0.89$. 
\begin{table}[hb]
\begin{center}
\begin{tabular}{ | m{3cm} | m{1cm}| m{1cm} | m{1cm} | } 
  \hline
  Events & $\rho_{\rm{BBH}}$ & $\rho_{\rm{res}}$ & $\mathfrak{F}$ \\
  \hline
  GW190519\_153544 & 15.34 & 6.38 & 0.92 \\
  \hline
  GW190828\_065509 & 9.67 & 6.30 & 0.84 \\
  \hline
  GW190706\_222641 & 13.39 & 7.80 & 0.86 \\ 
  \hline
  GW190708\_232457 & 13.97 & 6.00 & 0.92 \\ 
  \hline
\end{tabular}
\caption{Subset of events from O3 catalog that were followed up in residuals test show the residual SNR does not necessarily show any trend with network SNR recovered by a particular BBH waveform (cf. sec. IV A in~\cite{LIGOScientific:2020tif}). The fitting factor is calculated using Eq.~(\ref{eq:residual})}
\label{tab:residual}
\end{center}
\end{table}
 For $C=0.1$ panel in Fig.~\ref{fig:detectability} the black contour line encloses the region that will lead to sufficiently high residual SNR to indicate an exotic origin for the merger signal provided the signal has at least an SNR of $\sqrt{8^2+\rho_{\rm{BBH}}^2}$. Consistent with expectations, lower values of recovered $\rho_{\rm{BBH}}$ for a similar residual would make the test sensitive to a wider region of ECO mass in the $q_{\rm{ECO}}$--$M_{\rm{ECO}}$ parameter space. This can be easily spotted in table ~\ref{tab:residual} where we have shown a selection of events from the O3 catalog~\cite{Abbott_2023} for which residual test results were reported~\cite{LIGOScientific:2020tif}. While GW190519\_153544 and GW190828\_065509 have similar residual SNR, the event with the lower SNR has significantly lower $\mathfrak{F}$. On the other hand GW190706\_222641 and GW190708\_232457 have similar recovered SNR, but significantly different residual SNR, with the higher one corresponding to a lower $\mathfrak{F}$. The residuals for all of the events reported in~\cite{LIGOScientific:2020tif} were found to be consistent with noise, however, it is worth noting that this test is currently performed with the motivation of finding deviations from GR. Therefore, unless recovered SNRs and residuals show a correlation across all events, residuals in individual events are ruled out as noise artefacts. On the other hand if the catalogs comprise true BBH mergers and a set of ECO mergers mischaracterised as BBH mergers, we would expect only a few to have significant residuals at arbitrary recovered SNRs. Given the significantly high residual left behind by GW190828\_065509 despite the low recovered SNR (in comparison to other events of similar residual), it may be interesting to test further with exotic waveforms.

Naturally, if events detected with SNR above threshold leave enough residual SNR, this test would be able to distinguish the BBH hypothesis from a non-BBH hypothesis. We note here that both~\dtap~ and ~\deco~ lead to the same conclusions on detectability, with~\dtap~ giving slightly poorer fitting factors due to sharper drop in amplitude (cf. Figs.~\ref{fig:dtapamp} and~\ref{fig:decoamp}).

\subsection{Distinguishability}
%
The detectability analysis makes it clear that if stellar mass ECOs exist, then our current waveform models are well equipped to detect them except in the most extreme cases. The ability to characterize them as ECOs presents a challenge, particularly at low masses. We discussed one plausible indicator of exotic origin of the merger based on a residual analysis. We can also discuss distinguishability based on parameter inference. Using a chi-squared distribution for the three dimensional space of $q-M-\chi_{\rm{eff}}$, we can write the match-SNR correspondence as~\cite{Baird:2012cu}
\be
(1-\mathcal{M}) \geq \frac{3.1}{\rho_{\rm{dist}}^2} ,
\label{eq:distingsnr}
\ee
where $\mathcal{M}$ is the phase and time optimized match between~\dtap~ and~\bbh~. This means for the restricted subspace of  $q-M-\chi_{\rm{eff}}$, a template that matches 97\% of the signal would correspond to the boundary of $90\%$ credible interval (CI) at an SNR of 10. But for signals louder than $\rho>10$ a template of match 0.97 does not lie within the $90\%$ CI. Therefore, by randomly drawing from the $q-M-\chi_{\rm{eff}}$ posteriors it will be impossible to generate a waveform with~\bbh~that has a match of greater than 0.97 with the signal, if it was produced by~\dtap~. This would be a strong indicator of the signal being generated by exotic objects. 
\begin{figure}[h]
\includegraphics[width=0.49\textwidth]{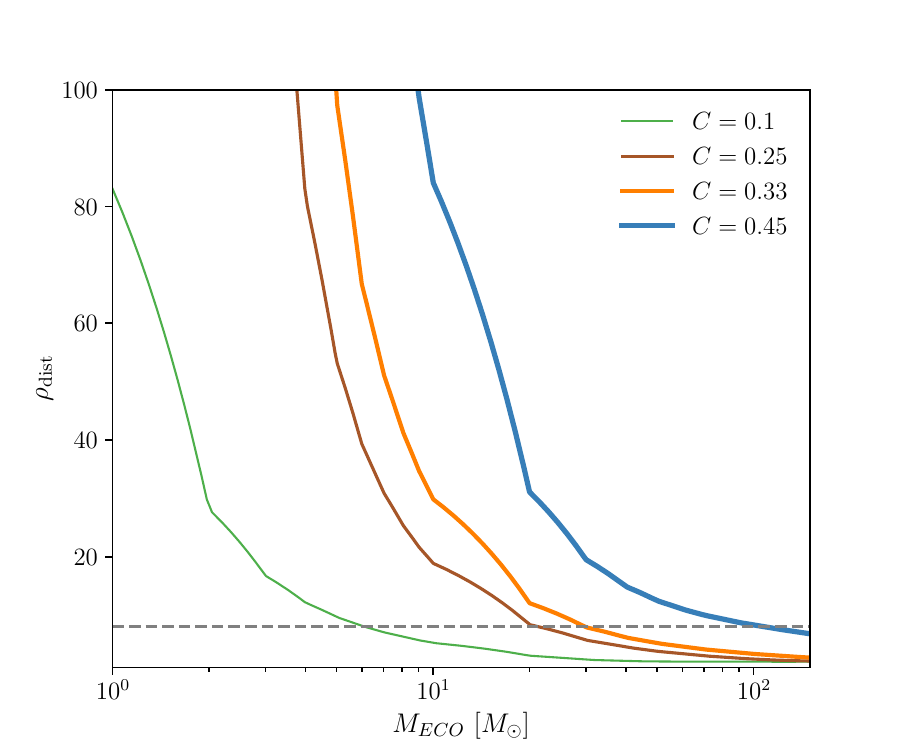}
\caption{ The minimum detectable SNR that enables distinguishing an ECO merger signal from a BBH merger with matches obtained using~\dtap. The gray dashed line shows the detectability threshold of 8. For masses $\gtrsim35\msun$ the distinguishability SNR is always lower than the detectability threshold for $C\leq0.33$. }
\label{fig:distinguishability}
\end{figure}

{ \ }
Inverting the above equation, we can compute the SNR at which an ECO merger would be distinguishable from a BBH merger. Figure~\ref{fig:distinguishability} shows the distinguishability SNR for different compactness as function of ECO total mass. It is easy to note from this plot that for low mass systems ($<10\msun$) and moderate to large compactnesses, it would be almost impossible to distinguish between an ECO merger and a BBH merger. On the other hand for masses $>10\msun$ distinguishability requires only moderately high SNR for $C\lesssim0.33$. In fact at masses $\gtrsim30\msun$ the distinguishability SNR is always lower than the detectability threshold of 8, suggesting that detected events will always be distinct. It is worthwhile to draw attention to an important caveat of our analysis at this point again; we ignore phase modulations due to finite-sized effects as well as tidal disruptions, and using a model specifically calibrated to such dephasing may significantly change distinguishability. Unfortunately this also means that using a model finely tuned to a particular model of ECO binary is as precarious as using a BBH model, in terms of distinguishability. In other words, unless a generalized framework is used to look for exotic objects, we are more likely to find what we are looking for! We discuss our first attempt in this direction in section~\ref{sec:compactness}. 

%
%
\begin{figure}
\includegraphics[width=\linewidth]{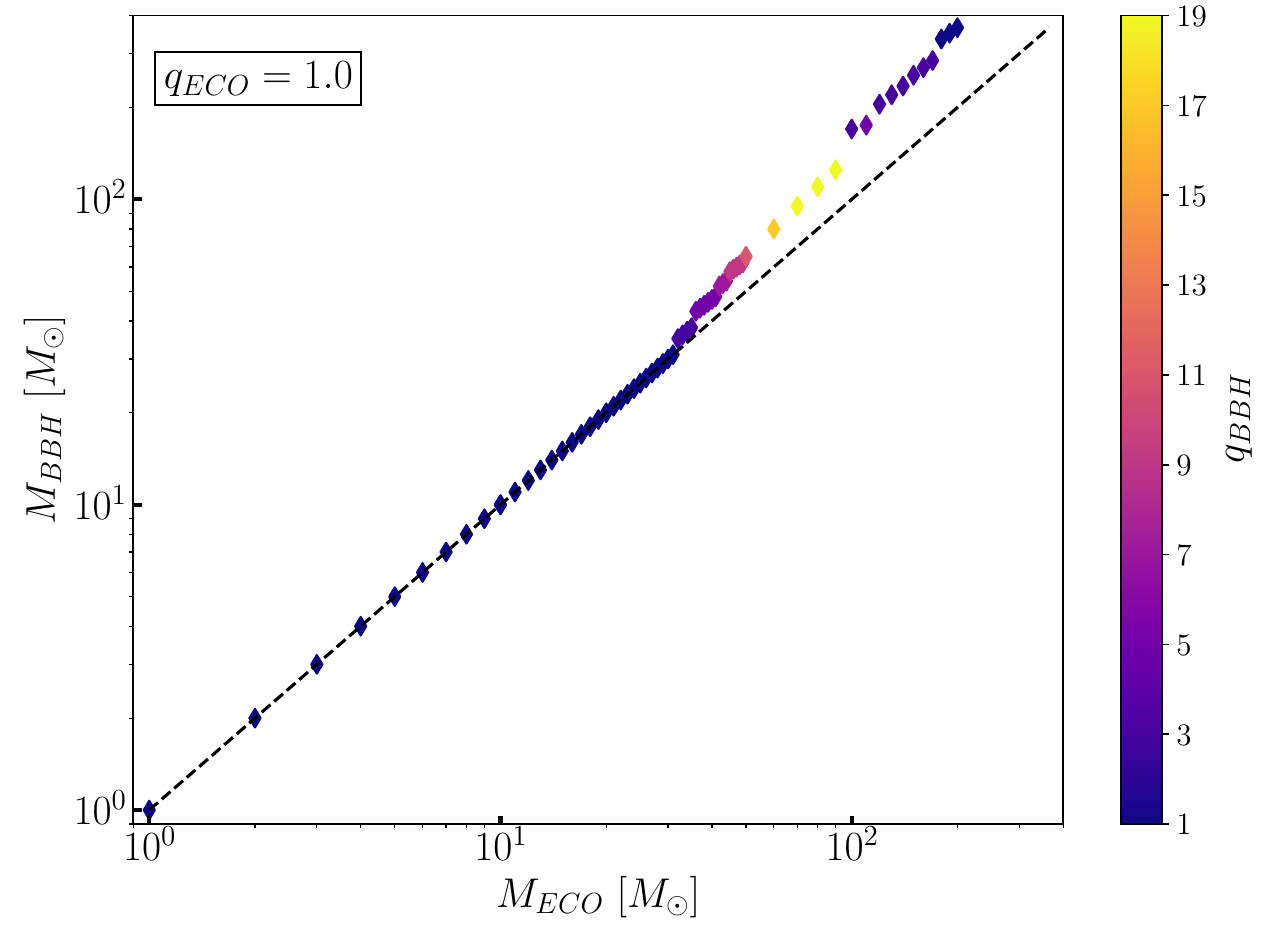}
\caption{Total mass and mass ratio recovery for equal mass ECO binaries using~\dtap~ as signal and ~\bbh~ as template; points on the black dashed line indicate no bias in total mass while the color of the marker represents the mass ratio of the best fit template. Biases in total mass can go upto 30\%~, but mass ratio biases can go upto 1800\%~(eg. for 100\msun) indicating higher mass ratio binaries are more likely to be ECO binaries mischaracterised as BBHs. }
\label{fig:parameterbiases}
\end{figure}
\subsection {Source parameters }


The fitting factor obtained by optimization across the three-dimensional parameter space in $q-M_{\rm{BBH}}-\chi_{\rm{eff}}$ gives a simplistic estimation of source parameters. For accurate recovery of source parameters Bayesian inference techniques need to be employed. Nevertheless, some qualitative observations can be made about the nature of biases, based on this restricted, grid-based analysis. 

Figure~\ref{fig:parameterbiases} gives an insight into the extent of biases in the mass parameters. For equal mass ECO signals of total mass $\gtrsim30$, the template prefers a higher total mass and mass ratio $q_{\rm{BBH}}>1$. To understand the interplay between mass ratio and total mass we again refer to the top panels of Fig.~\ref{fig:ampprofile}. While a higher total mass of the template captures the early truncation of the amplitude, it also increases the overall scaling of the amplitude envelope. The template tries to compensate for this by preferring a higher mass ratio, hence the $M-q$ biases are correlated. 

The above treatment of the biases has focussed only on the mass parameters. However, it is well known that the mass ratio and spin presents a degeneracy in the behaviour of the amplitude of the waveform. Furthermore, due to the well-known orbital hangup effect, it is reasonable to expect that adding anti-aligned spin to the template can mimic the effect of the inspiral ending early. However, the impact of optimizing over spin does not show monotonic variation with increasing (anti)aligned spin content and does not show a correlation with either of mass or mass ratio biases. Based on these biases we conclude that binaries that have been measured with BBH models to have unequal masses, especially with $M\gtrsim30\msun$, are the most likely candidates for further investigation. 


\section{ An ECO identifier}
\label{sec:compactness}
The distinguishability study done with ~\dtap~in Fig.~\ref{fig:distinguishability} shows that for loud enough events we should be able to tell if the ECO hypothesis is stronger than the BBH hypothesis. However, with a model that smoothly approaches the BBH limit as $C\rightarrow0.5$, it may even be possible to directly read off the ``exoticness'' of a binary from the inference on compactness. In this section we discuss the robustness of~\deco~in inferring compactness and the efficacy of compactness as an ECO identifier.

Being built upon~\bbh~, ~\deco~naturally inherits its regime of validity in generating reliable waveforms generation. Specifically, configurations beyond $q=18$ and $\chi_{\rm{eff}}=0.85$ (0.98 for equal mass binaries) fall outside calibration region and the model should be used with caution in this region. Additionally, we have tested the model in the range $C>0.05 $ and $q<20$, and do not see signs of pathological behaviour. The present version of the model is not valid beyond the Kerr limit.

Parameter estimation in GW data analysis typically involves updating the knowledge on the probability distribution of parameters that describe a signal using Bayes' theorem. The posterior probability distribution of the parameters is calculated as
\be
P(\theta|d) = \frac{P(d|\theta)P(\theta)}{P(d)},
\ee
where $P(\theta)$ is the prior probability distribution of the parameters $\theta$ and $P(d)$ is the total probability of the data consider all possible values of the parameters(or evidence). $P(d|\theta)$ is the likelihood function that quantifies the probability of the data given the parameters $\theta$. To facilitate effective source characterisation, a high dimensional parameter space is sampled by maximizing the multi-dimensional likelihood function, given the data, to obtain the joint posterior probability distribution for the parameters that best describe the data. ~\deco~ adds an extra parameter, compactness, to \textsc{Bilby}-- the standard bayesian framework of astrophysical inference used in GW data analysis~\cite{bilby}.

\subsection{Compactness}
As shown in Fig.~\ref{fig:decoamp}, the frequency domain amplitude of~\deco~ approaches that of a BBH merger for $C=0.5$. Therefore, we first want to verify that compactness estimates obtained using ~\deco~ are reliable. We generate a BBH signal with a total mass of $40\msun$ at $q=1$ and $\chi_{\rm{eff}}=0.1$ using~\deco~ and inject it at an SNR of $\sim25$ into stationary Gaussian noise at design sensitivity of the advanced LIGO detectors~\cite{adLIGOnoise} with zero detector noise. We attempt to recover it with ~\deco~ by sampling over compactness in addition to the intrinsic parameters of chirp mass ($\mathcal{M}_c$), mass ratio ($q$), aligned spin components ($\chi_{1,2}$), coalescence phase ($\phi_c$), and the extrinsic parameters of distance ($d_L$), sky location ($\alpha, \delta$), polarization ($\psi$) \& inclination ($\iota$). To ensure robustness of the compactness estimate we assigned a uniform distribution of prior probability in the range $\in [0.1,0.99]$. Note that, a compactness $>0.5$ is indicative of a Kerr BH.
\begin{figure}[h]
\includegraphics[width=0.49\textwidth]{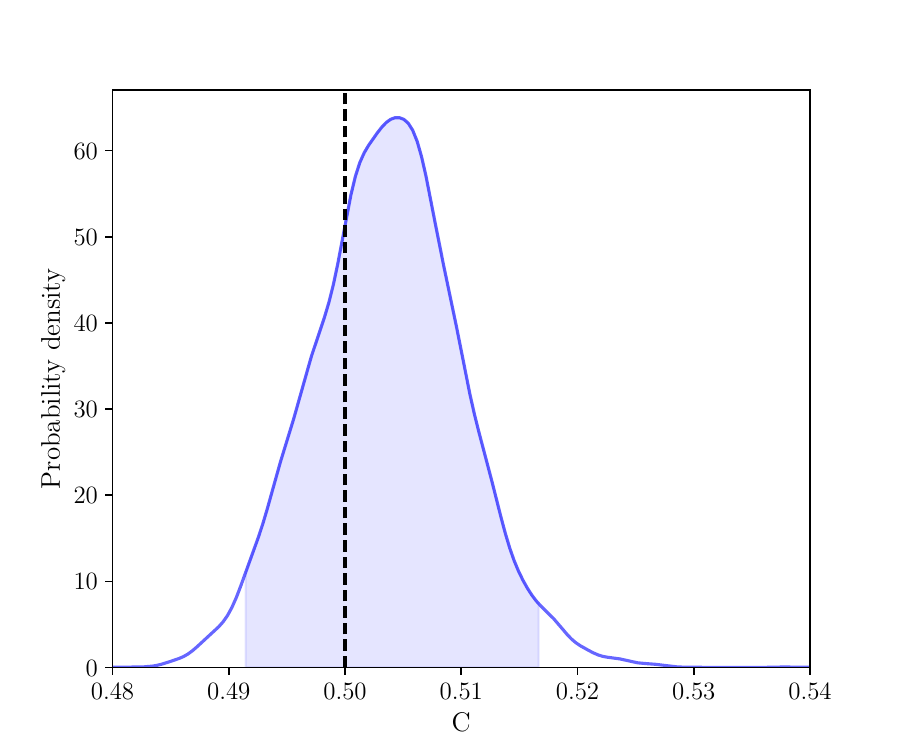}
\caption{Posterior probability distribution of compactness obtained using PhenomDECO as the recovery waveform, for an injected waveform generated with PhenomDECO at C=0.5}
\label{fig:comp5}
\end{figure}
\begin{figure}[h]
\includegraphics[width=0.49\textwidth]{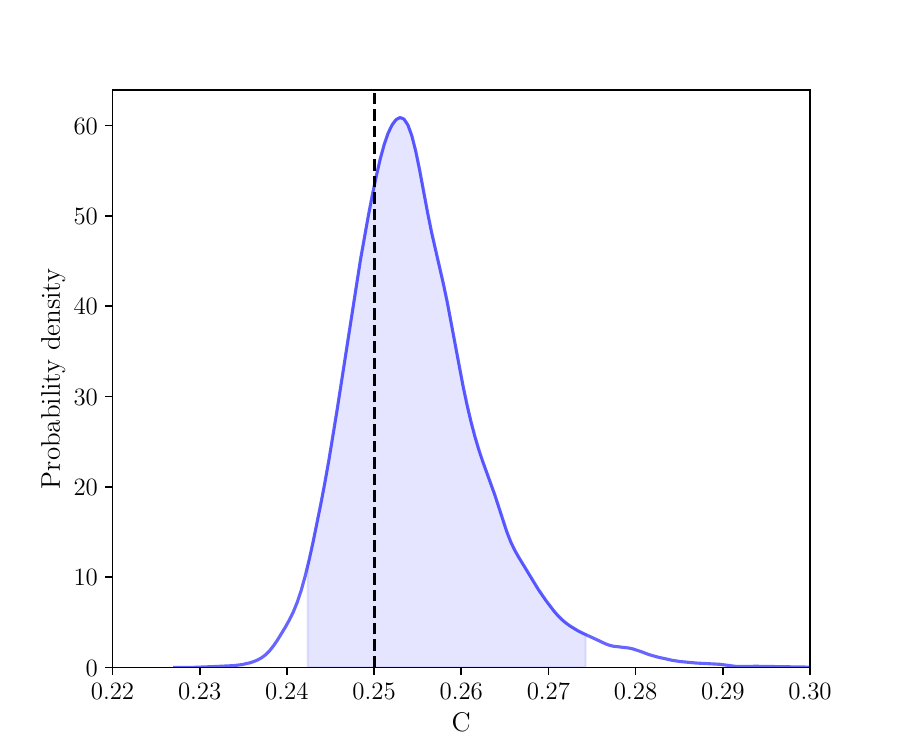}
\caption{Posterior probability distribution of compactness obtained using PhenomDECO as the recovery waveform, for an injected waveform generated with PhenomDECO at C=0.25}
\label{fig:comp2p5}
\end{figure}

Figure.~\ref{fig:comp5} shows the posterior probability distribution of compactness, while the injected value is shown by the black dashed vertical line. As is evident, the compactness is recovered with an accuracy $\sim1\%$. We note that despite the complexity of the high dimensional parameter space, and the use of a broad and uninformative prior, based on the simple earlier-merger morphology the compactness can be measured well even at moderate SNRs. We note a slight bias in the mass ratio recovery but $\mathcal{M}_c$ and $\chi_{\rm{eff}}$ are both recovered very well.

Now that we have established that this set up is indeed able to measure compactness, we evaluate the efficacy of the framework as an ECO identifier  with a~\deco~ injection at a compactness different than that of a Schwarzschild BH i.e., $C=0.25$, with all other intrinsic parameters the same as before. Figure.~\ref{fig:comp2p5} shows that compactness is yet again recovered with $\sim1\%$ accuracy; the recovery for $\mathcal{M}_c$, q and $\chi_{\rm{eff}}$ show similar trends as seen for C=0.5. 
This highlights that if nature truly produces ECO binaries that merge earlier than their BH counterparts due to a lower compactness, then ~\deco~ will not only be able to detect it but also identify it as distinct from a BBH merger. 
%

\subsection{Compactness estimates of real GW events}

\begin{figure*}[t]
\includegraphics[width=\textwidth]{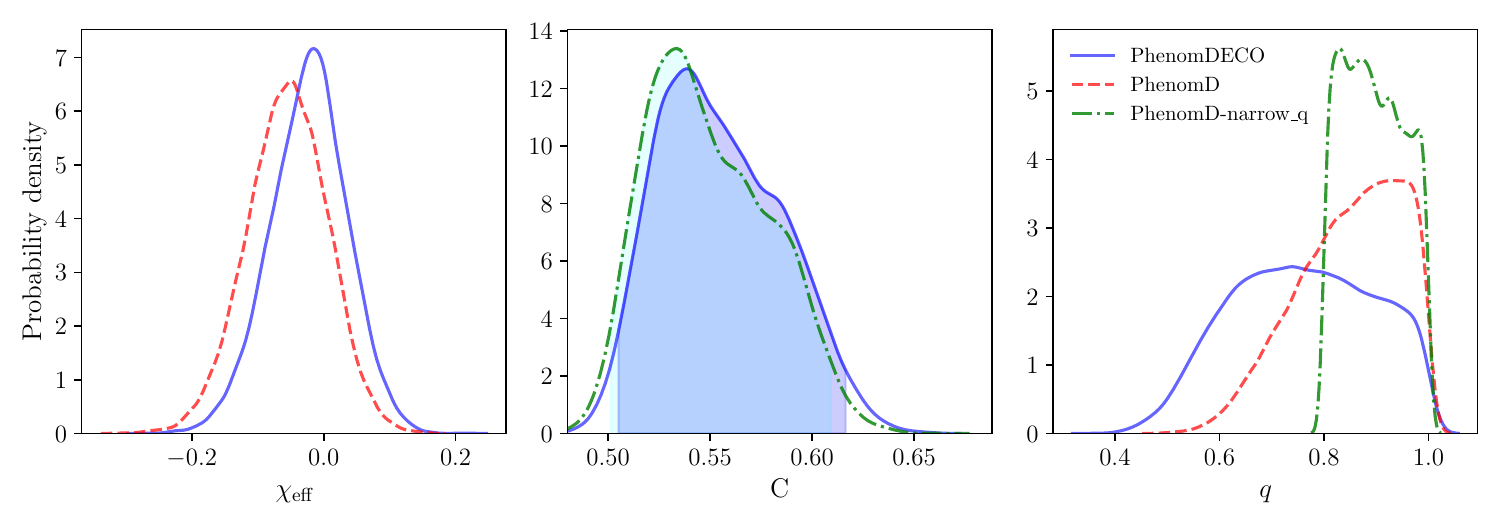}
\caption{Marginalised 1D posterior probability density of GW150914 \textit{left}:effective spin \textit{center}: compactness parameter and \textit{right}: mass ratio obtained with~\deco~ shown by blue, solid line. The probability densities with a standard BBH model~\bbh~ are shown by red, dashed line. The estimates with a prior probability on mass ratio restricted to [0.8,0.1] is shown by green, dot-dashed line. We show the 95\%~CI for only compactness parameter.}
\label{fig:gw150914comp}
\end{figure*}

\begin{figure}[h]
\includegraphics[width=0.49\textwidth]{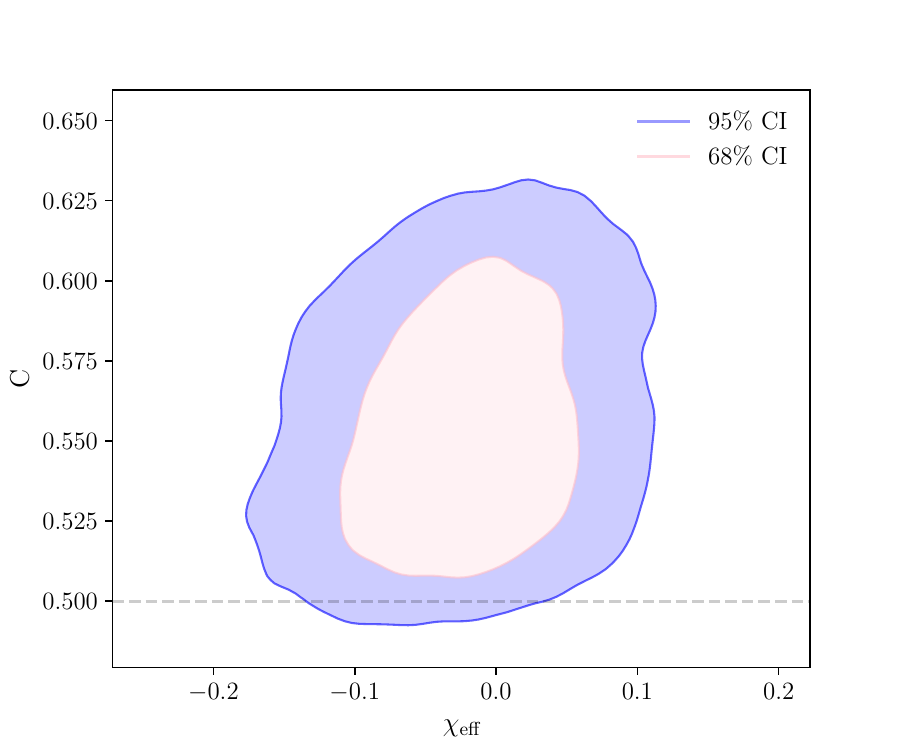}
\caption{Marginalised two-dimensional posterior probability distribution of compactness C and effective spin $\chi_{\rm{eff}}$ for GW150914 using~\deco~.}
\label{fig:gw150914_C_chi_2d}
\end{figure}

While the injections above demonstrate the significance compactness has on the signal, the robustness of the compactness measure can only be verified by estimates from real GW events. We performed Bayesian parameter estimation on the LIGO Hanford and Livingston data from the GW150914 observation~\cite{LIGOScientific:2016aoc} to infer compactness. The center panel on fig.~\ref{fig:gw150914comp} shows that~\deco~is able to 1) find a clear peak for the posterior probability distribution of compactness and 2) constrain $95\%$ of posterior probability density for compactness (i.e., $95\%$ credible interval) within $[0.505,0.616]$. Nominally, a compactness of $C=0.5$ is excluded at $95\% CI$, however, GW150914 has some support for aligned spin content as well as mass asymmetry in the binary as shown by the left and right panels of fig.~\ref{fig:gw150914comp} respectively. We know that a Kerr BH has a higher compactness than a Schwarzschild BH, although the two-dimensional posterior in Fig.~\ref{fig:gw150914_C_chi_2d} does not show a strong correlation.
This figure also reveals that while the 1D projection of the joint probability distribution excludes 0.5 at 95\% CI, it is actually included in the $C-\chi_{\rm{eff}}$ two-dimensional marginalized posterior probability distribution. We do not find  noticeable biases in recovery for any of the other parameters.

To confirm if the preference for higher mass ratio by ~\deco~ as seen in the right panel in fig.~\ref{fig:gw150914comp} can explain the broad posteriors for compactness and particularly a preference for significantly high median value, we used a narrower prior for $q\in[0.8,1]$ inspired by the q-posteriors from~\bbh~ and find that the compactness recovery with this prior is still very similar to that obtained with a uniform prior on $q$. This indicates that the compactness parameter does introduce some systematic biases in the mass ratio. We will explore further study of the $q-C-\chi_{\rm{eff}}$ biases in future work. Note also that the compactness posterior is not a sophisticated measure of the compactness of each individual components in a binary. All that it tells us is the closest approach between the two objects before the amplitude peaks, because~\deco~ inherently admits the same compactness for each object. Refinement in compactness estimates could also be explored by adapting~\deco~ to sample over compactness of individual objects.

\section{Conclusions} 
All GW observations to date have been from the orbits and merger of compact objects,  which makes binaries of exotic compact objects (i.e., neither black holes nor neutron stars) a particularly promising candidate for new discoveries in the future. However, if such sources exist, we are faced with three problems: (1) in many cases the signal may be indistinguishable from that produced by black-hole or neutron-star binaries, (2) if we are to accurately model the signal as we have done for black-hole binaries, then we will require a large quantity of computationally expensive NR simulations that cover the large (and perhaps infinite) parameter space of every potential source that we have imagined; and (3) such a modelling programme is oblivious to exotic sources that we \emph{have not} imagined. 

An alternative approach is to phenomenologically model signals that are agnostic to any particular theoretical model, and
instead include a range of broad, generic physical features. Such an approach would allow us to identify those physical 
features that will be most easily measurable, which has two advantages over targeting a narrow range of specific sources:
it allows us to identify the classes of objects that are most easily identifiable, thus guiding the direction of more precise
simulation and modelling efforts, and also to identify those physical features that can be measured independently of a
physical model, and are therefore more practical to target in observations. 

In this work we consider one simple example: the physical compactness of an object. We adapt a current BBH
waveform model to include a single simple phenomenological effect: if the objects are less compact than black holes, then
during their inspiral they will come into contact and merge earlier (i.e., at lower orbital frequency) than their BH
counterparts. This will cause the GW signal to terminate at a lower frequency. We illustrate the efficacy of this simple
modelling technique to identify when such signals might fail to be detected in current searches, and when, if observed, they
could be identified as exotic objects, and how well we can measure the compactness from this agnostic model. 

We find that for objects with $C \gtrsim 0.25$, i.e., up to twice as large as black holes, current BBH searches would rarely fail to detect observable signals for masses below 10\,$M_\odot$. Only for much lower compactness (i.e., objects 5 times larger than black holes) would we fail to detect most binaries with masses above 10\,$M_\odot$. However, the statement on detectability is incomplete without consideration of SNR distribution of the sources. Notably, the results in fig.~\ref{fig:detectability} can also be interpreted as the fractional loss in SNR in an observation, so that a merger signal of two $C=0.1$ ECOs and total mass of $30\msun$ with loud enough SNR (say 100) would be easily detectable by a~\bbh~ template, recovering an SNR$\sim50$.
 
This makes the fraction of the signal power that a BBH model cannot find in the data particularly interesting. For especially loud events, we can  translate these results into the residual signal power that will remain in the data after the best-matching BBH signal has been removed. A loud residual would tell us that we have detected a GW signal that is \emph{not} from a binary black hole, even without an ECO model to hand. 
If we \emph{do} have an ECO model, Fig.~\ref{fig:distinguishability} shows the SNR at which ECO signals will be distinguishable from BBHs.
For $C > 0.25$, signals with SNRs typical of current observations (less than $\sim$20) will not be distinguishable from BBH for masses below $\sim$10\,$M_\odot$. On the other hand, almost any objects with $C = 0.1$ will be easily distinguishable from BBH, so long as they can be detected.

Current residual tests are designed primarily to detect deviations from general relativity (GR). As a result, unless a correlation between recovered SNRs and residuals is observed across multiple events, residuals from individual events are typically interpreted as noise artefacts. But event catalogs may include both genuine BBH mergers and a subset of ECO mergers misclassified as BBH merger. In this context significantly large residuals, even if observed for a handful of events with no specific trend with recovered SNR, would be intriguing and may warrant further investigation using exotic waveform models.

We also consider the parameter biases for ECOs measured assuming they are BBH. We find that the mass ratio is biassed towards more asymmetric masses. This suggests that, when considering whether any current observations are ECOs, we should focus on those measured to have unequal masses. For instance, an event like GW190814~\cite{LIGOScientific:2020zkf}, where we observed a high mass ratio, could be an ECO, but based on fig.~\ref{fig:detectability} must have $C > 0.33$ since that the best-fitting BBH template has a fitting factor of $0.97$\cite{LIGOScientific:2020tif}.

Given that compactness has a significant effect on the signal, truncating the inspiral at lower frequencies, we might 
expect that compactness can be well measured in observations. Within the assumption of our limited signal model
(we have not included eccentricity, tidal deformation, or any attempt to capture the potential variations in  
phenomenology of the post-merger signal), we find that even for signals with only moderate SNRs ($\sim$25) it
is possible to measure the compactness to within $1\%$ accuracy. With this framework, we present the first compactness estimate of a real GW event, GW150914. Despite the very uninformative prior on the compactness, the posterior probability was found the be consistent with that of black hole. Note that ~\deco~  is built upon the assumption that both objects have the same compactness. We have also remarked on consolidating this measure to establish the BBH hypothesis for CBC events and will pursue constructing an astrophysically informed prior on compactness by running inference on the events from the CBC catalogs in future work.

These measurement results highlight a number of caveats to this work. We have included only \emph{one} 
physical effect in potential ECOs, namely a simple model of the effect of compactness on the contact point
that terminates inspiral. The model is entirely phenomenological and does not make additional checks on the physics of the source. However, the ability of~\deco~to smoothly sample over this new parameter and yielding consistent results with real data highlights the framework’s capacity to embed additional source physics within phenomenological models. The framework can be extended to include physics-based consistency checks, such as considerations for energy conservation due to the early end of inspiral, which we leave for future work. Furthermore, not all ECOs will follow this phenomenology. For example, consider one of the boson star simulations in Ref.~\cite{Evstafyeva:2024qvp} where, even with a compactness of $C=0.2$, the signal is almost identical to that from a BBH. Aside from ignoring tidal effects and post-merger dynamics, this is also partly because that work uses a different interpretation for compactness than the usual $M/R$ definition used for BHs. This reminds us that our results apply to ECOs where compactness impacts the signal in one specific way, but does not necessarily apply to all ECOs, or all definitions of compactness. 

However, we do not consider this a weakness of our model, but rather it highlights an important point: we should identify
physical properties that are measurable, and focus on those. If the compactness used in those simulations
could not be measured in an observation, then perhaps it is not the ideal property with which to parametrise
a class of signals. In studying signals, and modelling them, and determining how they can be used
in measurements, we need to identify which properties are most clearly measurable, and first model
their impact on signals before moving on to other effects. This is somewhat analogous to the situation in
binary-black-hole modelling, where the most easily measurable spin effect is the mass-weighted sum of
the aligned-spin components, $\chi_{\rm eff} = (m_1 \chi_1 + m_2 \chi_2)/M$~\cite{Ajith:2009bn}, rather than the individual
spin components $\chi_1$ and $\chi_2$.

%

Similarly, a strength of such an approach also is that it allows us to determine the properties of compact objects
that could be measured without making any prior assumptions about the nature of the object. By 
contrast, if one produces a model of signals from, for example, boson-star binaries, and then observes
a signal consistent with that model (and nott consistent with BBH), we cannot then conclude that the
objects are in fact boson stars: only that they are more like boson stars than black holes. They may 
in fact be the manifestation of some entirely different physical phenomena that leads to compact objects
that, when in binaries, produce GW signals more akin to boson-star binaries than black-hole binaries. 
A more agnostic approach would allow us to measure general properties of ECOs, and from there 
determine the range of physical processes that could produce such objects. Once we identify which physical properties have the largest influence on the observed signal morphology, we can prioritize these features in follow-up NR simulations. Therefore, compactness estimates from~\deco~can inform NR simulation priorities -- and of course, simulation results are necessary to refine~\deco~ as an ECO identifier.

\section*{Acknowledgements}

S.G was supported by the Max Planck Society’s Independent Research Group program. M.H. was supported by Science and Technology Facilities Council (STFC) grant ST/V00154X/1. S.G. is grateful to Charlie Hoy, Duncan MacLeod \& Frank Ohme for helpful suggestions on data analysis and to Katy Clough \& Tamara Evstafyeva for discussions on exotic compact objects. We thank Jonathan Thompson for useful comments. The authors acknowledge the computational resources provided by the LIGO laboratory and supported by National Science Foundation Grants PHY-0757058 and PHY-0823459, which were used to obtain the bayesian inference results presented in this paper.

This research has made use of data obtained from the Gravitational Wave Open Science Center (gwosc.org), a service of the LIGO Scientific Collaboration, the Virgo Collaboration, and KAGRA. This material is based upon work supported by NSF’s LIGO Laboratory which is a major facility fully funded by the National Science Foundation, as well as the Science and Technology Facilities Council (STFC) of the United Kingdom, the Max-Planck-Society (MPS), and the State of Niedersachsen/Germany for support of the construction of Advanced LIGO and construction and operation of the GEO600 detector. Additional support for Advanced LIGO was provided by the Australian Research Council. Virgo is funded, through the European Gravitational Observatory (EGO), by the French Centre National de Recherche Scientifique (CNRS), the Italian Istituto Nazionale di Fisica Nucleare (INFN) and the Dutch Nikhef, with contributions by institutions from Belgium, Germany, Greece, Hungary, Ireland, Japan, Monaco, Poland, Portugal, Spain. KAGRA is supported by Ministry of Education, Culture, Sports, Science and Technology (MEXT), Japan Society for the Promotion of Science (JSPS) in Japan; National Research Foundation (NRF) and Ministry of Science and ICT (MSIT) in Korea; Academia Sinica (AS) and National Science and Technology Council (NSTC) in Taiwan.

Various plots and analyses in this paper were made using software packages \texttt{LALSuite}~\cite{lalsuite},  \texttt{Matplotlib}~\cite{Hunter:2007}, \texttt{Numpy}~\cite{Harris_2020}, \texttt{Scipy}~\cite{2020SciPy-NMeth} and \texttt{Seaborn}~\cite{Waskom2021}.

%

\bibliography{references_arxiv.bib}
\end{document}